\documentclass{article}
\usepackage{amssymb}
\usepackage{amsmath}
\def\1ad{\mbox{\normalsize $^1$}}
\def\2ad{\mbox{\normalsize $^2$}}
\def\3ad{\mbox{\normalsize $^3$}}
\def\4ad{\mbox{\normalsize $^4$}}
\def\5ad{\mbox{\normalsize $^5$}}
\def\6ad{\mbox{\normalsize $^6$}}
\def\7ad{\mbox{\normalsize $^7$}}
\def\8ad{\mbox{\normalsize $^8$}}
\def\makefront{
\vspace*{1cm}\begin{center}
\def\sp{
\renewcommand{\thefootnote}{\fnsymbol{footnote}}
\footnote[4]{corresponding author : \email_speaker}
\renewcommand{\thefootnote}{\arabic{footnote}}
}
\def\newtitleline{\\ \vskip 5pt}
{\Large\bf\titleline}\\
\vskip 1truecm
{\large\bf\authors}\\
\vskip 5truemm
\addresses
\end{center}
\vskip 1truecm
{\bf Abstract:}
\abstracttext
\vskip 1truecm
}
\numberwithin{equation}{section}
\def\cg{{\mathfrak g}}  
\def\cN{{\cal N}}

\def\Bc{{\mathbb C}} 
\def\Br{{\mathbb R}} 
   
\def\eps{\epsilon}  
\def\im{{\rm Im}}
\def\re{{\rm Re}} 
\begin {document}                 
\def\titleline{
%
%
%
Nontrivial RR two-form field strength and     %
SU(3)-structure                               %
}
\def\email_speaker{
{\tt 
%
%
peter.kaste@cpht.polytechnique.fr
}}
\def\authors{
%
%
%
%
%
Peter Kaste, Ruben Minasian, Michela Petrini and  Alessandro 
Tomasiello
}
\def\addresses{
%
%
%
%
Centre de Physique Th{\'e}orique, Ecole 
Polytechnique\footnote{Unit{\'e} mixte du CNRS et de l'EP, UMR 7644}\\ 
91128 Palaiseau Cedex, France\\
\mbox{} \\
kaste, ruben, petrini, tomasiel@cpht.polytechnique.fr
}
\def\abstracttext{
%
%
We discuss how in the presence of a nontrivial RR two-form field
strength and nontrivial dilaton
the conditions of preserving supersymmetry 
on six-dimensional manifolds lead to  
generalized monopole and Killing spinor equations. 
We show that the manifold is
K\"ahler in the ten-dimensional string frame if $F_0^{(1,1)}=0$.
We then determine explicitly the intrinsic torsion of the
SU(3)-structure on six-manifolds that result via 
Kaluza-Klein reduction from seven-manifolds with
G$_2$-structure of generic intrinsic torsion.
Lastly we give explicitly the
intrinsic torsion of the SU(3)-structure for an $\cN$=1 supersymmetric
background in the presence of nontrivial RR
two-form field strength and nontrivial dilaton.
}
\large
\makefront
\section{Introduction}
%
 
A better understanding of $\cN$=1 supersymmetric compactifications of
string theory to four dimensions is an important step towards more
realistic string theories. A promising avenue to take is to break the
$\cN$=2 supersymmetry of the well-studied Calabi-Yau compactifications
of type II string theories down to $\cN$=1 by including a background
of RR field strength that may describe either internal RR fluxes or
spacetime filling D-branes. Of course their presence back-reacts on the
metric. In particular the new
supersymmetric ground state is no longer a Calabi-Yau manifold. It is
an interesting question to study the geometry of these minimally
supersymmetric ground states and to characterize how their structure
deviates from the one of a Calabi-Yau. 

In \cite{kmpt.02} these questions have been addressed for
compactifications of type IIA with a background of nontrivial RR
two-form field strength and nontrivial dilaton. This is the situation
that is easiest to analyze, since the triple $(g,F,\varphi)$ of the
groundstate metric $g$, the background two-form field strength $F$ and
the dilaton $\varphi$ can be described in terms
of a G$_2$ manifold $Y$. Namely this is the internal manifold of the
purely geometrical M-theory compactification which via Kaluza-Klein
reduction gives rise to the above type IIA configuration.

In these proceedings we first 
construct explicitly an SU(3)-structure 
$(g_X,J,\psi_3)$ on the six-dimen\-sio\-nal base
space $X$ obtained by Kaluza-Klein reduction from a seven-dimensional
manifold $Y$ with G$_2$-structure. We then analyze 
the constraints that preserving $\cN$=1 supersymmetry in four dimensions
imposes on this SU(3)-structure by requiring the G$_2$-structure on
$Y$ to be torsion-free. These constraints can be cast into a primitivity
constraint on $F$, a monopole equation relating $d\varphi$ to $F$ and
a Killing spinor equation on the associated SU(3)-invariant spinor. 
It immediately follows that the manifold $(X,g_X,J)$ is K\"ahler 
in the string frame if
the primitive part of $F^{(1,1)}$ vanishes.

Next we compute explicitly the intrinsic torsion of the SU(3)-structure
$(g_X,J,\psi_3)$ on $X$ for a generic G$_2$-structure on $Y$. Specializing
this result to the case of torsion-free G$_2$-structure, we give the
intrinsic torsion of the SU(3)-structure for an $\cN$=1 supersymmetric
background in the presence of nontrivial RR
two-form field strength and nontrivial dilaton. 
Concretely,
we show that
in the notation of \cite{CS} its components are given by
$W_1=W_2^-=W_3=0$, $W_2^+=-\check{F}_0^{(1,1)}$, 
$W_4=-(\beta-2\alpha) d\varphi$ and 
$W_5=-(\beta-3\alpha) d\varphi$.
 
This intrinsic torsion is the obstruction for the Levi-Civita
connection of $(X,g_X)$ to have holonomy SU(3). It can therefore be
seen as a measure of how the manifold fails to be a Calabi-Yau. 
Recently the concept of intrinsic torsion of $G$-structures has been
applied to compactifications with background fields in
\cite{gaun.02a,louis.02,luest.02,gaun.02b,gold.02}. 
In \cite{louis.02} it is
in particular argued to describe the mirror of NS three-form fluxes
in an otherwise purely geometrical mirror compactification. 
For $\cN$=1 supersymmetric compactifications to three dimensions in
the presence of background fields $F$ and $\varphi$ and for further
references we refer to \cite{kmpt.02}.
                                           
\section{From a G$_2$-structure to an SU(3)-structure}

Let $Y$ be the seven-dimensional manifold 
to which one lifts in M-theory and
$g_Y$ the background metric on $Y$. 
Via Kaluza-Klein reduction it is related to the
background metric $g_X$ on the internal space $X$ of the type IIA
compactification by
\begin{equation}
ds_Y^2 = e^{-2\alpha\varphi} ds_X^2+ 
 e^{2\beta\varphi} (dz+A)^2 , \label{gygx}
\end{equation}
where $A$ is the RR one-form potential and $\varphi$ the dilaton of
type IIA. The parameters $\alpha$ and $\beta$ determine the frame of
$g_X$ in ten-dimensional type IIA. 
They take the values $(\alpha,\beta)$=$(1/3,2/3)$ for 
the string frame.

Suppose $Y$ carries a G$_2$-structure. 
The latter is specified by the doublet
$(g_Y,\Phi)$ where $\Phi$ is a G$_2$-invariant, nowhere vanishing
three-form on $Y$. It can be represented as
\begin{equation}
\Phi=\frac{1}{3!} \phi_{ABC} \hat{e}^A\hat{e}^B\hat{e}^C , \label{Phi}
\end{equation}
where $\hat{e}^A$ for $A=1,\ldots,7$ is a frame of orthonormal one-forms
w.r.t.\ $g_Y$ and $\phi_{ABC}$ are the structure constants of the
imaginary octonions. Moreover, the G$_2$-structure singles out a
unique G$_2$-invariant spinor $\epsilon$. It is real (Majorana) 
and satisfies
\begin{equation}
\gamma_{AB}\, \eps = i\,\phi_{ABC} \,\gamma ^C \eps \qquad \mbox{and} \qquad
\phi_{ABC} = -i\eps^{\dagger} \gamma_{ABC} \eps. \label{7dga} 
\end{equation}
If and only if the G$_2$-structure is torsion-free, the Levi-Civita
connection associated to $g_Y$ has holonomy in $G_2$ and in that case
the spinor $\eps$ will be covariantly constant w.r.t.\ the Levi-Civita
connection.    
This spinor $\eps$ on $Y$ is then the internal part of the supersymmetry
generator in the $\cN$=1 supersymmetric M-theory compactification on
$Y$. 
It is projectible onto $X$ along the U(1)-fibers of the
Kaluza-Klein bundle $\pi : Y\mapsto X$ if it is constant along these
fibers. In that case it becomes the internal part of the
supersymmetry generator in type IIA. 
Note that since this constant
along the fiber may vary over $X$, the spinor $\eps$ as seen on $X$
has a U(1) gauge symmetry. In particular $X$ need only carry a
Spin$^c$-structure and not a Spin-structure.  

If $\eps$ is projectible and
whether the G$_2$-structure is torsion-free or not,
the identity (\ref{7dga}) projects on $X$ onto
\begin{equation} 
\gamma_{ab}\eps = i\,\psi_{abc} \gamma ^c \eps + i\,J_{ab} \gamma \,\eps 
\qquad \mbox{and} \qquad 
\gamma_a \gamma \,\eps  = -i\,J_{ab} \gamma ^b \eps ,  
\label{6dga} 
\end{equation} 
where we have defined $\psi_{abc}\equiv\phi_{abc}$ and 
$J_{ab}\equiv\phi_{ab7}$
for $a,b,c=1,\ldots,6$ and where $\gamma=\gamma^7$ is the chirality
operator on $X$. Furthermore
\begin{equation}
\psi_{abc} = -i \eps^{\dagger} \gamma_{abc} \eps \qquad \mbox{and} \qquad
J_{ab} = -i \eps^{\dagger} \gamma_{ab}\gamma \eps . \label{6dgb}
\end{equation}
The spinor $\eps$ on $X$ can hence be used to build nowhere vanishing
forms 
\begin{equation}
\psi_3 \equiv \frac{1}{3!} \psi_{abc} e^a e^b e^c \qquad \mbox{and}
\qquad 
J \equiv  \frac{1}{2} J_{ab} e^a e^b 
\end{equation}
on $X$. Moreover, 
since $J_{a}^{~b}J_{b}^{~c}=-\delta_{a}^{~c}$, this $J_a^{~b}$  
defines a natural almost complex structure on $X$ with respect to
which $g_X$ is automatically hermitian, with associated two-form
$J$. Splitting each tangent plane into a holomorphic and
antiholomorphic space w.r.t. $J_a^{~b}$ one sees that $\psi_3$ is the
real part of a $(3,0)$-form $\Omega=\psi_3-i (*\psi_3)$. In particular
the forms $\psi_3$ and $J$ are not only nowhere vanishing on $X$, they
are also by construction invariant under the action of SU(3) on the
tangent bundle $TX$. Altogether, the triple $(g_X,J,\psi_3)$ defines
an SU(3)-structure on $X$. Its associated SU(3) invariant spinor is
$\eps$. 

In summary, if $Y$ is a (Kaluza-Klein) U(1)-bundle over $X$ and has
a G$_2$-structure with an associated projectible G$_2$-invariant
spinor, then the base $X$ carries an SU(3)-structure. This implies
that the structure group of $TX$ is SU(3). However, the
SU(3)-structure on $X$ will generically have torsion, even if the
G$_2$-structure was torsion-free. 
I.e.\ although
there exist connections on $TX$ that are compatible with the metric
and have holonomy SU(3), generically none of them will be torsion-free. In
that case the Levi-Civita connection cannot have holonomy SU(3) and
$X$ is not a Calabi-Yau. The intrinsic torsion of the SU(3)-structure
$(X,g_X,J,\psi_3)$ is the obstruction for it to be a Calabi-Yau.

In order to see how $X$ fails to be a Calabi-Yau, we look at the
differential equations satisfied by $\eps$ on $X$ or alternatively by
$\psi_3$ and $J$.

\section{The monopole and Killing spinor equations} 
   
Let the G$_2$-structure on $Y$ be torsion-free so that we preserve
$\cN$=1 supersymmetry in four dimensions.
The covariant constancy of $\eps$ on $Y$ reduces to the following
system on $X$
\begin{subequations} 
\begin{eqnarray}
\left( D_a + \frac{i}{2}\, \alpha\, (\partial_b \varphi)J_{~a}^{b}
  \,\gamma\right) \eps +  
i\, \left( \frac{1}{2}\, \alpha\, (\partial_b \varphi)\, \psi ^b_{\ ac}  
- \frac {1}{4} \check{F}_{ab} J^b_{\ c}\right) \gamma ^c \eps &=& 0 , 
\label{6dgrav}\\ 
\left( \frac{1}{4} \check{F}_{ab}J^{ab}\right) \gamma \eps +   
\left( \frac{1}{4} \check{F}_{ab}\psi^{ab}_{\ \ c}  
-\, \beta \,(\partial_a \varphi) J^{a}_{\ c}\right) \gamma ^c \eps &=& 0 , 
\label{6ddil} 
\end{eqnarray}
\end{subequations}
where we have defined 
$\check{F}\equiv e^{(\alpha+\beta)\varphi}F$ and where $D_a$ denotes
the covariant derivative w.r.t.\ the Levi-Civita connection on $(X,g_X)$. 
Since $\gamma^A\eps$ are
linearly independent, the latter of these equations gives
\begin{subequations} 
\begin{align}
F^{ab}J_{ab} &= 0 \quad & \Leftrightarrow & \quad 
& J\lrcorner \, F &= 0 , \label{monopole.1} \\ 
\beta\, (\partial_a \varphi)J^a_{~c} &= \frac{1}{4}\check{F}^{ab}\psi_{abc} 
\quad & \Leftrightarrow & \quad  
& \beta\, d\varphi &= \frac{1}{2} \check{F}\lrcorner \, (*\psi_3) ,
\label{monopole.2}
\end{align}
\end{subequations} 
where $\lrcorner$ denotes the contraction of forms w.r.t.\ the metric
$g_X$.
On a K\"ahler manifold the first of these would imply that
$F$ is primitive and even though $(X,g_X,J)$ might not be K\"ahler we
will refer to (\ref{monopole.1}) as a primitivity constraint. 
The second equation is a generalized monopole
equation. It relates $d\varphi$ to the $(2,0)$ and $(0,2)$ parts of
$F$ w.r.t.\ the almost complex structure. 
Inserting these into (\ref{6dgrav}) leads to the following Killing
spinor equation on $X$ for $\beta=2\alpha$,
\begin{equation}
\left( D_a + \frac{i}{2}\, \alpha\, (\partial_b \varphi)J_{~a}^{b} \,\gamma
 -\frac{i}{8} \left[\check{F}_{ab}J^b_{~c}+\check{F}_{cb}J^b_{~a}
 \right] \gamma ^c \right) \eps =0 . \label{6dgrav.b}
\end{equation} 
This implies that the only nonvanishing components of the Nijenhuis
tensor associated to the almost complex structure 
$J_a^{~b}=-i\eps^{\dagger}\gamma_a^{~b}\gamma\eps$ are
\begin{equation}
N^{\bar{a}}_{bc} = \frac{i}{2} 
 \left( \check{F}_{c\bar{d}}\, \epsilon^{\bar{d}~\bar{a}}_{~b}
       -\check{F}_{b\bar{d}}\, \epsilon^{\bar{d}~\bar{a}}_{~c} \right)
\qquad \mbox{and} \qquad
N^{a}_{\bar{b}\bar{c}} = -\frac{i}{2} 
 \left( \check{F}_{\bar{c}d}\, \epsilon^{d~a}_{~\bar{b}}
       -\check{F}_{\bar{b}d}\, \epsilon^{d~a}_{~\bar{c}} \right)
     , \label{Nijenhuis.hol}
\end{equation}
where we have used the holomorphic/antiholomorphic basis of $TX$
w.r.t.\ $J_a^{~b}$. The almost complex structure defined by the spinor
$\eps$ is therefore integrable if and only if $F^{(1,1)}\equiv
0$. Since $dJ=0$ we find that in this case $(X,g_X,J)$ is
K\"ahler for $\beta=2\alpha$, i.e. defines a torsion-free U(3)-structure. 
The Killing spinor equation on $X$ then reduces to
\begin{equation} 
\left( D_a + \frac{i}{2}\, \alpha\, (\partial_b \varphi)J_{~a}^{b}
 \,\gamma\right) \eps  
=0  .
\end{equation}
Since $\eps$ is not covariantly constant w.r.t.\ the Levi-Civita
connection for nontrivial $\varphi$, $F^{(2,0)}$ and $F^{(0,2)}$, 
the SU(3)-structure $(X,g_X,J,\psi_3)$ however still has torsion. 

%
\section{The general relation between the intrinsic G$_2$-torsion and
  SU(3)-torsion}

The intrinsic torsion of the G$_2$-structure $(g_Y,\Phi)$ 
takes values in $(\cg_2)_{\perp}\otimes T^*Y$, where 
so(7)=$\cg_2\oplus(\cg_2)_{\perp}$ and can be decomposed as \cite{CS}
\begin{equation}
X_1 \in Y\otimes \Br , \qquad
X_2 \in \Lambda^2_{14} T^*Y , \qquad
X_3 \in \Lambda^4_{27} T^*Y , \qquad
X_4 \in \Lambda_{7} T^*Y , \label{def.Xs}
\end{equation} 
where $\Lambda^n_m T^*Y$ denotes $n$-forms that transform in the
representation $\underline{m}$ of G$_2$. In particular these
representations imply that
\begin{equation}
X_2 \lrcorner_{g_Y} \Phi =0 , \qquad 
\Phi \lrcorner_{g_Y} X_3 =0 , \qquad
(*\Phi) \lrcorner_{g_Y} X_3 =0 , \label{X.constraints}
\end{equation} 
where $\lrcorner_{g_Y}$ denotes the contraction of forms w.r.t. the
metric $g_Y$.
The components (\ref{def.Xs})
are determined through $d\Phi$ and $d(*\Phi)$ as \cite{CS}
\begin{subequations} 
\begin{eqnarray}
d\Phi &=& X_1 (*\Phi) + X_4 \wedge \Phi + X_3 , \label{dPhi.X} \\
d(*\Phi) &=& \frac{4}{3} X_4 \wedge (*\Phi) + X_2 \wedge \Phi
\label{dsPhi.X} . 
\end{eqnarray}
\end{subequations}
Let us furthermore introduce the notation
\begin{equation}
X_j = Y_j + Z_j \wedge \hat{e}^7 , \quad \mbox{for}~~j=1,\ldots,4.
\end{equation}
Then the constraints (\ref{X.constraints}) take the form
\begin{subequations} 
\begin{eqnarray}
X_2 \lrcorner_{g_Y} \Phi &=0 \quad \Leftrightarrow &
\left\{ \begin{array}{r@{~=~}l}
e^{\alpha\varphi}Y_2 \lrcorner_{g_X} \psi_3 + Z_2 \lrcorner_{g_X}  J & 0 , \\
J \lrcorner_{g_X} Y_2 & 0 , \label{X2.constraint} \\
\end{array} \right. \\
\Phi \lrcorner_{g_Y} X_3 &=0  \quad \Leftrightarrow &
\left\{ \begin{array}{r@{~=~}l}
e^{\alpha\varphi}\psi_3 \lrcorner_{g_X} Y_3 - J \lrcorner_{g_X}  Z_3 & 0 , \\
\psi_3 \lrcorner_{g_X} Z_3 & 0 , \label{X3a.constraint} \\
\end{array} \right. \\
(*\Phi) \lrcorner_{g_Y} X_3 &=0 \quad \Leftrightarrow &
\frac{1}{2} e^{\alpha\varphi} J^2 \lrcorner_{g_X} Y_3
 + (*\psi_3) \lrcorner_{g_X}  Z_3 = 0 . \label{X3b.constraint}
\end{eqnarray}
\end{subequations}
In the following all the contractions will be taken w.r.t.\ the metric
$g_X$ and we will drop the label, i.e.\ 
$\lrcorner\equiv\lrcorner_{g_X}$. 

Analogously the intrinsic torsion of the SU(3)-structure $(g_X,J,\psi_3)$ 
takes values in su(3)$_{\perp}\otimes T^*X$, where 
so(6)=su(3)$\oplus$su(3)$_{\perp}$. It can be decomposed as
\begin{align}
W_1 &\in X \otimes \Bc , \quad &
W_2 &\in \Lambda^{(1,1)}_{8\oplus 8} T^*X |_{\rm primitive}
 , \nonumber \\
W_3 &\in (\Lambda^{(2,1)}_{6} T^*X \oplus \Lambda^{(1,2)}_{\bar{6}}
 T^*X)|_{\rm primitive} , \quad &
W_4 &\in \Lambda^{(1,0)}_{3} T^*X \oplus \Lambda^{(0,1)}_{\bar{3}}
 T^*X , \label{def.Ws} \\
W_5 &\in \Lambda^{(1,0)}_{3} T^*X \oplus \Lambda^{(0,1)}_{\bar{3}}
 T^*X , & & & \nonumber
\end{align}
where $\Lambda^{(n_1,n_2)}_m T^*X$ denotes $(n_1,n_2)$-forms that
transform in the 
representation $\underline{m}$ of SU(3). In particular these
representations imply that
\begin{equation}
J\lrcorner \, W_2 = 0 \quad , \quad 
J\lrcorner \, W_3 = 0 \quad , \quad 
\psi_3 \lrcorner \, W_3 = 0 \quad \mbox{and} \quad
(*\psi_3) \lrcorner \, W_3 = 0 . \label{W.constraint}
\end{equation}
We can similarly 
express the components of its intrinsic torsion through $dJ$,
$d\psi_3$ and $d(*\psi_3)$ as \cite{CS}
\begin{subequations} 
\begin{eqnarray}
dJ &=& \frac{3}{2} \im(\overline{W_1} \Omega) + W_4 \wedge J + W_3 
 \label{dJ.W} \\
 &=& \frac{3}{2} W_1^- \psi_3 - \frac{3}{2} W_1^+ (*\psi_3) 
     + W_4 \wedge J + W_3 , \nonumber \\
d\psi_3 &=& W_1^+ J^2 + W_2^+ \wedge J + \re(W_5 \wedge
     \overline{\Omega}) \label{dpsi3.W} \\ 
 &=& W_1^+ J^2 + W_2^+ \wedge J + W_5^+ \wedge \psi_3 
     + W_5^- \wedge (*\psi_3) , \nonumber \\
d(*\psi_3) &=& W_1^- J^2 + W_2^- \wedge J + \im(W_5 \wedge
 \overline{\Omega}) \label{dspsi3.W} \\  
 &=& W_1^- J^2 + W_2^- \wedge J + W_5^+ \wedge (*\psi_3) 
     - W_5^- \wedge \psi_3 , \nonumber \\
d\Omega &=& W_1 J^2 + W_2 \wedge J + \overline{W_5}\wedge \Omega
     \label{dOm.W} , 
\end{eqnarray}
\end{subequations}
where we have used the (3,0)-form
\begin{equation}
\Omega = \psi_3 -i(*\psi_3) , 
\end{equation} 
as well as
\begin{equation}
W_j = W_j^+ -i W_j^- ,~~\mbox{for}~~j=1,2,5 .
\end{equation}  
The minus signs on the imaginary parts result from the fact that
compared to \cite{CS} we use the opposite
orientation, where $J=e^{14}+e^{25}+e^{36}$. 

Using 
\begin{subequations} 
\begin{eqnarray}
\Phi &=& e^{-3\alpha\varphi}\psi + e^{-2\alpha\varphi} J\wedge \hat{e}^7 , \\
(*\Phi) &=& e^{-4\alpha\varphi}(*J) + e^{-3\alpha\varphi} 
 (*\psi_3)\wedge \hat{e}^7
 = -\frac{1}{2} e^{-4\alpha\varphi}J^2 + e^{-3\alpha\varphi} 
 (*\psi_3)\wedge \hat{e}^7 ,
\end{eqnarray}
\end{subequations}
equations (\ref{dPhi.X}), (\ref{dsPhi.X}) and
(\ref{dJ.W})--(\ref{dOm.W}), as well as 
$\check{F}= e^{(\alpha+\beta)\varphi}F$,
one derives the following four identities,
\begin{subequations} 
\begin{eqnarray}
\lefteqn{3\alpha\, \psi_3 \wedge d\varphi + W_1^+ J^2 + W_2^+ \wedge J
 + \re(W_5 \wedge \overline{\Omega}) + J \wedge \check{F}} \nonumber \\
&=&  
-\frac{1}{2} e^{\alpha\varphi}X_1 J^2 -\psi_3\wedge Y_4 +
 e^{3\alpha\varphi} Y_3 , \label{dPhi.n7} \\
\lefteqn{(\beta-2\alpha) J\wedge d\varphi + \frac{3}{2} W_1^- \psi_3 
 - \frac{3}{2} W_1^+ (*\psi_3) +W_4 \wedge J +W_3} \nonumber \\
&=&
e^{-(\alpha+\beta)\varphi}X_1 (*\psi_3) + e^{-\beta\varphi}J\wedge Y_4 
 - e^{-(\alpha+\beta)\varphi}Z_4 \psi_3
 + e^{-(\beta-2\alpha)\varphi} Z_3 , \label{dPhi.w7} \\ 
\lefteqn{2\alpha\, J^2\wedge d\varphi - J^2 \wedge W_4 - (*\psi_3)\wedge
 \check{F} } \nonumber \\
&=&
-\frac{2}{3} J^2\wedge Y_4 + e^{\alpha\varphi} \psi_3\wedge Y_2 ,
 \label{dsPhi.n7} \\ 
\lefteqn{(\beta-3\alpha) d\varphi \wedge (*\psi_3) + W_1^- J^2 
+ W_2^- \wedge J + \im(W_5 \wedge \overline{\Omega})} \nonumber \\
&=&  
-\frac{4}{3} e^{-\beta\varphi}(*\psi_3)\wedge Y_4 
-\frac{2}{3} e^{-(\alpha+\beta)\varphi}Z_4 J^2
+e^{(\alpha-\beta)\varphi}J\wedge Y_2
+e^{-\beta\varphi} \psi_3 \wedge Z_2  . \label{dsPhi.w7}
\end{eqnarray}
\end{subequations}
From these equations we can project onto the various $W_j$'s by
suitable contractions and express them through the intrinsic torsion
$(Y_j,Z_j)$ of the G$_2$-structure. 
To this end it is convenient to decompose forms as
\begin{subequations} 
\begin{eqnarray}
F&=& F^{(0)}\, J + F_0^{(1,1)} + F^{(2,0)} + F^{(0,2)} \label{split.F} \\
&=& \frac{1}{3} (J\lrcorner \, F)J + F_0^{(1,1)}
 + \frac{1}{8} \left[ (F\lrcorner \, 
  \overline{\Omega} )\lrcorner \, \Omega + (F\lrcorner \,
  \Omega )\lrcorner \, \overline{\Omega}\, \right] , \nonumber \\ 
Y_2&=& Y_2^{(0)}\, J + Y_{2,0}^{(1,1)} + Y_2^{(2,0)} + Y_2^{(0,2)}
\label{split.Y2} \\ 
&=& \frac{1}{3} (J\lrcorner \, Y_2)J + Y_{2,0}^{(1,1)}
 + \frac{1}{8} \left[ (Y_2 \lrcorner \,
  \overline{\Omega} )\lrcorner \, \Omega + (Y_2 \lrcorner \,
  \Omega )\lrcorner \, \overline{\Omega}\, \right] , \nonumber \\ 
Z_3 &=& Z_3^{(0)} \psi_3 + \tilde{Z}_3^{(0)} (*\psi_3)
 + (Z_3^{(1,0)}+Z_3^{(0,1)})\wedge J + Z_{3,0}^{(2,1)}
 +Z_{3,0}^{(1,2)} \label{split.Z3} \\
&=& \frac{1}{4} (\psi_3 \lrcorner \, Z_3) \psi_3
 + \frac{1}{4} ((*\psi_3) \lrcorner \, Z_3) (*\psi_3)
 + \frac{1}{2} (J \lrcorner \, Z_3)\wedge J + Z_{3,0}^{(2,1)}
 +Z_{3,0}^{(1,2)} , \nonumber \\
Y_3 &=& Y_3^{(0,1)} \wedge \Omega + Y_3^{(1,0)} \wedge \overline{\Omega}
 + (Y_3^{(2,0)}+Y_3^{(0,2)})\wedge J + Y_{3,0}^{(2,2)}
 +Y_{3,0}^{(1,1)} \wedge J + Y_3^{(0)} J^2 , \nonumber \\
\label{split.Y3}
\end{eqnarray}
\end{subequations}
where subscripts ``0'' denote primitive forms.
Using projectors such as
\begin{equation}
Y_{3,0}^{(1,1)} = J\lrcorner \, Y_3 -\frac{1}{3} (J^2 \lrcorner \, Y_3) J
 -\frac{i}{8} (\Omega \lrcorner \, Y_3) \lrcorner \, \overline{\Omega} 
 +\frac{i}{8} (\overline{\Omega} \lrcorner \, Y_3) \lrcorner \, \Omega ,
\end{equation} 
the components $W_j$ of the intrinsic torsion of the SU(3)-structure
$(g_X,J,\psi_3)$ can be expressed through the components 
$(Y_j,Z_j)$ of the intrinsic torsion of the G$_2$-structure
$(g_Y,\Phi)$ as,
\begin{subequations} 
\begin{eqnarray}
W_1^+ &=& -\frac{2}{3} e^{-(\alpha+\beta)\varphi}X_1 
 -\frac{2}{3} e^{-(\beta-2\alpha)\varphi} \tilde{Z}_3^{(0)} ,
 \label{W1p} \\
W_1^+ + \check{F}^{(0)} 
&=& -\frac{1}{2} e^{\alpha\varphi}X_1 
 + e^{3\alpha\varphi} Y_3^{(0)} ,\label{W1p.f} \\
W_1^- &=& -\frac{2}{3} e^{-(\alpha+\beta)\varphi}Z_4 
 , \\
W_2^+ 
+ \check{F}_0^{(1,1)} 
&=& e^{3\alpha\varphi} Y_{3,0}^{(1,1)} , \label{W2p.f}  \\
W_2^- 
&=& e^{(\alpha-\beta)\varphi} Y_{2,0}^{(1,1)} , \\
W_3 &=& e^{-(\beta-2\alpha)\varphi} [ Z_{3,0}^{(2,1)}
 +Z_{3,0}^{(1,2)} ] , \\
W_4 + (\beta-2\alpha) d\varphi 
&=& e^{-\beta\varphi} Y_4 + e^{-(\beta-2\alpha)\varphi}
 [ Z_3^{(1,0)} + Z_3^{(0,1)} ] , \label{W4} \\ 
W_4 - 2\alpha\, d\varphi +\frac{1}{2} \check{F}\lrcorner \, (*\psi_3) 
&=& \frac{2}{3} Y_4 - \frac{1}{2} e^{\alpha\varphi}
Y_2\lrcorner \, \psi_3 , \label{W4.f} \\ 
W_5^{(1,0)} - 3\alpha\, d\varphi^{(1,0)} 
 +\frac{i}{4} \check{F}\lrcorner \, \Omega 
&=& Y_4^{(1,0)} - \frac{1}{4} e^{3\alpha\varphi}
\Omega \lrcorner \, Y_3 , \label{W5.f} \\ 
W_5^{(1,0)} + (\beta- 3\alpha) d\varphi^{(1,0)} 
&=& \frac{4}{3} e^{-\beta\varphi} Y_4^{(1,0)}
 - \frac{1}{4} e^{(\alpha-\beta)\varphi} Y_2 \lrcorner \, \Omega
 -i e^{-\beta\varphi} Z_2^{(1,0)} . \label{W5} 
\end{eqnarray}
\end{subequations}
The decomposition of the $X_j$'s into the
$W_j$'s at the level of representations
figures already in \cite{CS}. Equations 
(\ref{W1p})--(\ref{W5}) determine the explicit coefficients that
appear in this decomposition for a metric (\ref{gygx}) of the form
that appears in Kaluza-Klein reductions to an arbitrary frame.

Let's now specialize again to the case of torsion-free
G$_2$-structure, where all the right hand sides vanish. From
(\ref{W1p}) and (\ref{W1p.f}) we recover the primitivity constraint
(\ref{monopole.1})
\begin{equation}
\frac{1}{3} J\lrcorner \, F \equiv F^{(0)} = 0 , \label{prim}
\end{equation}
whereas equations (\ref{W4}), (\ref{W4.f}) as well as 
(\ref{W5.f}),(\ref{W5}) and their complex conjugates give us the
monopole equation (\ref{monopole.2})
\begin{equation}
\beta\, d\varphi = \frac{1}{2} \check{F} \lrcorner \, (*\psi_3)
. \label{monop} 
\end{equation}
In addition to these two constraints, the components of the intrinsic
torsion of the SU(3)-structure in the case of torsion-free
G$_2$-structure are given by
\begin{align}
W_1^{\pm} &=0 \quad , \quad & W_2^- &=0 \quad , \quad 
& W_3 &= 0 , \nonumber \\
W_2^+ &= -\check{F}_0^{(1,1)} \quad , \quad
& W_4 &= -(\beta-2\alpha) d\varphi \quad , \quad
& W_5 &= -(\beta-3\alpha) d\varphi . 
\end{align}
Moreover, we recover the result from the previous section that 
for $\beta=2\alpha$ the manifold $(X,g_X,J)$ is
K\"ahler if $F^{(1,1)}_0=0$, since then the only nonvanishing component 
of the intrinsic torsion of $(g_X,J,\psi_3)$ is $W_5$.
\\ \mbox{} \\

{\bf Acknowledgement} \\
We would like to thank B.~Acharya, D.~Calderbank, M.~Douglas, 
D.~Martelli and A.~Moroianu for useful discussions. 
PK would furthermore like to thank the organizers of the 35th
Ahrenshoop Symposium for the invitation to participate in a
stimulating conference and the opportunity to present this work.  
This work is supported in part by EU contract HPRN-CT-2000-00122 and by 
INTAS contracts 55-1-590 and 00-0334. PK and MP are supported by 
European Commission Marie 
Curie Postdoctoral Fellowships under contract numbers  
HPMF-CT-2000-00919 and HPMF-CT-2001-01277.  


\begin{thebibliography}{77}

\bibitem{kmpt.02}
P.~Kaste, R.~Minasian, M.~Petrini and A.~Tomasiello, 
``Kaluza-Klein bundles and manifolds of exceptional holonomy,''
JHEP {\bf 0209} (2002) 033, arXiv:hep-th/0206213. 

\bibitem{CS}  
S.~Chiossi, S.~Salamon, 
``The intrinsic torsion of $SU(3)$ and $G_2$ structures'' 
Proc. conf. Differential Geometry Valencia 2001, math.DG/0202282. 

\bibitem{gaun.02a}
J.P.~Gauntlett, D.~Martelli, S.~Pakis and D.~Waldram,
``G-structures and wrapped NS5-branes,'' arXiv:hep-th/0205050.

\bibitem{louis.02}
S.~Gurrieri, J.~Louis, A.~Micu and D.~Waldram,
``Mirror symmetry in generalized Calabi-Yau compactifications,''
arXiv:hep-th/0211102.

\bibitem{luest.02}
G.L.~Cardoso, G.~Curio, G.~Dall'Agata, D.~L\"ust, P.~Manousselis,
G.~Zoupanos, ``Non-K\"ahler string backgrounds and their five torsion
classes,'' arXiv:hep-th/0211118.

\bibitem{gaun.02b}
J.P.~Gauntlett and S.~Pakis, ``The geometry of d=11 Killing spinors,''
arXiv:hep-th/0212008. 

\bibitem{gold.02}
E.~Goldstein and S.~Prokushkin, ``Geometric model for complex
non-K\"ahler manifolds with SU(3)-structure,''
arXiv:hep-th/0212307. 

\bibitem{cglp} 
M.~Cvetic, G.~W.~Gibbons, H.~Lu and C.~N.~Pope,  
``Almost special holonomy in type IIA and M theory,''  
Nucl.\ Phys.\ {\bf B 638} (2002) 186,  
arXiv:hep-th/0203060.  

\end{thebibliography}
\end{document}